\documentclass[12pt, a4paper]{article}
\usepackage{moreverb,bm,xspace}
\usepackage[colorlinks,bookmarksopen,bookmarksnumbered,citecolor=red,urlcolor=red]{hyperref}
\usepackage{tikz}
\newcommand\BibTeX{{\rmfamily B\kern-.05em \textsc{i\kern-.025em b}\kern-.08em T\kern-.1667em\lower.7ex\hbox{E}\kern-.125emX}}
\newcommand{\ie}{{\em i.e.\/}\xspace}
\newcommand{\eg}{{\em e.g.\/}\xspace}

\usepackage[margin=1in]{geometry}  
\usepackage{graphicx}              
\usepackage{amsmath}               
\usepackage{amsfonts}              
\usepackage{amsthm}                
\usepackage[T1]{fontenc}
\usepackage[utf8]{inputenc}
\usepackage{authblk}
\usepackage{breqn}

\begin{document}

\title{Variable Selection in Covariate Dependent Random Partition Models: an Application to Urinary Tract Infection}

\author[1]{William Barcella\thanks{william.barcella.13@ucl.ac.uk}} 
\author[1]{Maria De Iorio}
\author[1]{Gianluca Baio}
\author[2]{James Malone-Lee}
\affil[1]{Department of Statistical Science, University College London, London, UK}
\affil[2]{University College London Medical School, London, UK}

\renewcommand\Authands{ and }

\maketitle
\begin{abstract}
Lower urinary tract symptoms (LUTS) can indicate the presence of urinary tract infection (UTI), a condition that if it becomes chronic requires expensive and time consuming care as well as leading to reduced quality of life. Detecting the presence and gravity of an infection from the earliest symptoms is then highly valuable. Typically, white blood cell count (WBC) measured in a sample of urine is used to assess UTI. We consider clinical data from 1341 patients at their first visit in which UTI (i.e. WBC$\geq 1$) is diagnosed. In addition, for each patient, a clinical profile of 34 symptoms was recorded. In this paper we propose a Bayesian nonparametric regression model based on the Dirichlet Process (DP) prior aimed at providing the clinicians with a meaningful clustering of the patients based on both the WBC (response variable) and possible patterns within the symptoms profiles (covariates). This is achieved by assuming a probability model for the symptoms as well as for the response variable. To identify the symptoms most associated to UTI, we specify a spike and slab base measure for the regression coefficients: this induces dependence of symptoms selection on cluster assignment. Posterior inference is performed through Markov Chain Monte Carlo methods.

\end{abstract}

\section{Introduction}\label{sec:intro}
In medical settings, individual level data are often collected for the relevant subjects on a variety of variables; these typically include background characteristics (\eg sex, age, social circumstances) as well as information directly related to the interventions being applied (\eg clinical measurements such as blood pressure or the results of a particular test). This set up applies for both experimental and observational studies --- perhaps even more so in the latter case, when data are often (although not always) collected using registries or administrative databases.  

Arguably, the most common use of such data involves some form of regression analysis where the main ``outcome'' variable is related to (some of) the covariates (or ``\textit{profiles}'') that have been collected. More specifically, clinicians may be interested in identifying suitable subgroups of patients presenting similar features; this categorisation can be used, for example, to suitably apply the optimal treatment for the (sub)population that will benefit the most. Alternatively, the focus may be on finding the covariates that best describe the variation in the outcome, for example in order to determine which symptoms should be measured to better characterise the chance that a new (as yet unobserved) patient is affected by a particular disease. The first of these tasks can be framed in the broader statistical problem of \textit{clustering}, while the second one is an example of \textit{model selection} (also called \textit{variable selection}). 

More interestingly, because complex and heterogeneous data are increasingly often collected and used for the analysis, a further connection between clustering and model selection can be considered, \ie  that one (set of) covariate(s) may be relevant in explaining the outcome variable for a subset of subjects, but not for others. In other words, the two tasks can be mixed in a more comprehensive analysis strategy to produce cluster-specific model selection.

For example, the dataset motivating this work contains records of Lower Urinary Tract Sympthoms (LUTS), a group of symptoms indicating dysfunctions of the lower urinary tract, \ie bladder and urethra, including incontinence and dysuria. This symptoms are related to several diagnosis: from anxiety, to multiple sclerosis or bladder tumour. However, the most common diagnosis associated with LUTS is Urinary Tract Infection (UTI). To examine possible UTI, samples of urine are analysed counting the number of White Blood Cells (WBC). The presence of WBC in urines, even in very small quantities, reveals UTI \cite{kupelianetal2013, dukes1928} and very often large number of WBC are associated with high degree of inflammation. The database includes LUTS and WBC for a number of patients affected by UTI (\ie WBC$\geq1$). For each individual, the set of LUTS constitute the patient's profile, while WBC can be considered as an indicator of UTI, the actual condition being investigated; the clinical objective is to assess the potential relationship between the symptoms and the infection. 

A standard approach to deal with these problems is to employ generalised linear models, including random effects (usually modelled using a Normal distribution) to account for heterogeneity between patients. This is clearly a restrictive assumption in many applications as often the distribution of the random effects is non-Normal, multi-modal, or perhaps skewed. In our analysis, we move beyond the traditional parametric hierarchical models, in order to account for the known patient heterogeneity that cannot be described in a simple parametric model. This heterogeneity is a common feature of many biomedical data and assuming a parametric distribution or mis-specifying the underlying distribution would impose unreasonable constraints; this in turn may produce poor estimates of parameters of interest. It is therefore important to use non-parametric approaches to allow random effects to be drawn from a sufficiently large class of distributions. That is the modelling strategy we adopt in this paper. 

In order to take into account the heterogeneity among the patients, it is convenient to study the relationship between the covariates and the response within groups of patients having similar symptoms profiles and similar levels of WBC (\ie in a clustering setting). In addition, it is crucial to evaluate which symptoms are explanatory of the level of WBC within each group (\ie in a variable selection setting). The goal is to develop a method for assessing the relationship between a response variable (in our case WBC) and a set of covariates (the profile) within clusters of patients with similar characteristics, in order to make prediction about the response for a new patients. This will ultimately provide valuable information on the mechanisms of action of the underlying disease being investigated.

To this aim, we develop a modelling strategy based on Bayesian non-parametric methods that allows us to accomplish both tasks at once. We propose a (potentially infinite) mixture of regression models to link the response with the covariates, where also the weights of the mixture can depend on the covariates. In this way, observations will be clustered based on the information contained in both the clinical profiles and the outcome variable. Within each cluster, variable selection is achieved employing \textit{spike and slab} prior distributions that assign positive probability to the regression coefficients being equal to zero. The Bayesian framework allows us to perform both tasks simultaneously in a probabilistically sound way, so that clustering and variable selection inform each other. The results of the application on LUTS data show that our formulation leads to improved predictions, in comparison to other existing methods. 

The rest of the work is organised as follows. In Sections \ref{litrev} and \ref{litrev2} we briefly describe  clustering and  variable selection problems from a Bayesian perspective and we review the most relevant literature. Then in Section \ref{sec:srpmx} we introduce the details of our proposed approach and briefly explain how to perform posterior inference while in Section \ref{sec:varsel} we show how to summarise posterior inference output in a meaningful way. In Section \ref{sec:luts} we present an application of our model to the LUTS dataset mentioned above. Finally, in Section \ref{sec:con} we discuss our results and draw conclusions.

\section{Clustering via Bayesian non-parametrics methods}\label{litrev} 
Suppose we wish to investigate the clustering structure characterising a dataset $\bm{y} = (y_1,...,y_n)$. In this case we may assume that the observations come from $K$ distinct clusters, each characterised by a parameter $\theta_k$ ($k=1,\ldots, K$), which specifies the probability distribution of the data in each cluster (note that, in general, $\theta_k$ can itself be a vector). This implies that the probability distribution of each observation $y_i$ is a mixture: 
\begin{equation}\label{eq:fmm}
y_i \mid\bm\theta, \bm\psi \sim \sum_{k=1}^K\psi_k p(y_i \mid \theta_k),
\end{equation}
where $\bm{\theta}=(\theta_1,\ldots,\theta_K)$ and $\bm\psi=(\psi_1,\ldots,\psi_K)$, with $\psi_k$ indicating the probability that unit $i$ belongs in cluster $k$. 

This approach is often referred to as ``model-based clustering'' and its aims include the estimation of $\psi_k$ and $\theta_k$, as well as of the number $K$, which is often the most challenging task. There are two popular approaches in the Bayesian literature to estimate the number of clusters: the first one consists in fitting separate models for different values of $K$ and then applying a model selection criterion (such as AIC, BIC or DIC) to identify the specification with most support from the data. The second one is fully grounded in a Bayesian framework and amounts to specifying a prior on $K$ and then performing full posterior inference on all the model parameters. The latter approach is adopted in this paper by employing a non-parametric prior to allow for more flexibility.

The model in (\ref{eq:fmm}) can be extended to an \textit{infinite} mixture model 
\begin{equation}\label{eq:infmm}
y_i \mid\bm\theta, \bm\psi \sim \sum_{k=1}^{\infty}\psi_k p(y_i\mid \theta_k)
\end{equation} 
--- notice that although the structure in (\ref{eq:infmm}) allows for an infinite number of clusters {\em a priori}, by necessity in a dataset made by $n$ observations there can be at most $n$ distinct groups. 
Within the Bayesian framework, an elegant and efficient way of defining this model is to select a Dirichlet Process (DP) prior \cite{ferguson1973}.

A DP is a stochastic process whose realisations are probability distributions such that if a random distribution $\mathcal{G}\sim\text{DP}(\alpha, \mathcal{G}_0)$ then it is almost surely discrete \cite{ferguson1973,blackwell1973,ohlssenetal2007}. $\mathcal{G}_0$ denotes the base measure (representing some sort of baseline distributions around which $\mathcal{G}$ is centred) and $\alpha \in\mathbb{R}^+$ is the precision parameter. DP models are by far the most widely used non-parametric Bayesian model, mainly because of computational simplicity; this derives from the fact that the complexity in simulating from the relevant posterior distributions is essentially dimension-independent. 

A ``constructive'' definition of a DP is given in \cite{sethuraman1994} using the following representation: 
\begin{displaymath}
\mathcal{G}=\sum_{k=1}^{\infty}\psi_k\delta_{\theta_k},
\end{displaymath}
where $\delta_{\theta_k}$ is the Dirac measure taking value 1 in correspondence of $\theta_k$ and 0 otherwise. The infinite set of parameters $\theta_1,\theta_2,\ldots$ is drawn independently from the continuous distribution $\mathcal{G}_0$ and the weights $\psi_1,\psi_2,\ldots$ are constructed using a \textit{stick breaking} procedure \cite{ishwaranetal2001}:
\begin{displaymath}
\psi_k = \phi_k\prod_{j=1}^{k-1}(1-\phi_j)
\end{displaymath}
with each $\phi_j \sim \text{Beta}(1, \alpha)$. The almost sure discreteness of the random distribution $\mathcal{G}$ is particularly relevant in the case of clustering, since $\mathcal{G}$ will provide the weights and the locations for the mixture in (\ref{eq:infmm}), leading to a Dirichlet Process Mixture (DPM) \cite{antoniak1974}, which can be rewritten in a compact way as 
\begin{eqnarray}
\bm{y}\mid \bm{\theta} & \sim & p(\bm{y}\mid\bm\theta) \nonumber \\
\bm{\theta}\mid\ \mathcal{G} & \sim & \mathcal{G} \label{eq:dpm} \\
\mathcal{G} & \sim & \text{DP}(\alpha, \mathcal{G}_0). \nonumber
\end{eqnarray}
For example, if $p( \bm{y} \mid \bm{\theta})$ is taken to be a Normal distribution with  $\bm\theta=(\bm\mu,\bm\sigma^2)$, then the distribution of $\bm{y}$  becomes an infinite mixture of Normal kernels with weights determined by the DP prior, \ie 
$\bm{y}\sim \sum_{k=1}^\infty \psi_k \mbox{ Normal}(\bm{y}\mid \mu_k,\sigma^2_k)$. 

Given its discreteness, setting a DP prior on the parameter vector $\bm{\theta}$ implies a non-zero probability that two or more of its elements are equal. This, in turn, implies that the DP imposes a clustering structure on the data so that the observations will be grouped together in $k\leq n$ clusters, each characterised by a specific distribution. The parameter $\bm{\psi}$ includes the prior probabilities of belonging in each cluster and $\theta_k$ denotes the cluster-specific parameter. The advantage of this strategy is that the number of components $k$ is also learned from the data through the posterior distribution. Thus, the vector of individual-level parameters $\theta_1,\ldots,\theta_n$ reduces to the vector of unique values $\theta_1^*,\ldots,\theta_k^*$ assigned to the $n$ observations.

This is also evident integrating out  $\mathcal{G}$ from the joint distribution of $(\theta_1,\ldots,\theta_n)$ and writing the conditional prior for each $\theta_i$ as in \cite{blackwelletal1973}:
\begin{equation}\label{eq:condprior}
\theta_i\mid\bm{\theta}_{(i)} \sim \frac{\alpha}{\alpha+n-1} \mathcal{G}_0 + \frac{1}{\alpha + n -1}\sum_{i'\neq i}\delta_{\theta_{i'}}
\end{equation}
where $\bm{\theta}_{(i)}$ is obtained removing the $i$-th component from $(\theta_1,\ldots,\theta_n)$. The discrete part in (\ref{eq:condprior}) induces ties among the components of $(\theta_1,\ldots,\theta_n)$. In particular, $\theta_i$ will take an already observed value within $\bm{\theta}_{(i)}$ with probability $1/(\alpha + n -1)$, while a new value (from $\mathcal{G}_0$) with probability $\alpha/(\alpha +n -1)$.

The latter links the DPM model of (\ref{eq:dpm}) with a Random Partition Model (RPM), \ie a probabilistic model over partitions of the $n$ observations, which are imposed by the DP prior. We denote with $\rho_n$ the partition of the $n$ observations in $k$ clusters, each characterised by the values of $\theta_j^*$ for $j=1,\ldots,k$. A membership vector $\bm{s}=(s_1,\ldots,s_n)$, describes which of the $\theta_j^*$'s is associated with each observation. Thus, $s_i$ takes on the values $\{1,2,\ldots,k\}$ for $i=1,\ldots,n$. From (\ref{eq:condprior}) and assuming $\theta_i$ as the last value to be sampled (this holds for every $\theta_i$ since (\ref{eq:condprior}) is exchangeable \cite{blackwelletal1973}), the prior distribution over all the possible partitions of $n$ observations implied by a DP is 
\begin{equation}\label{eq:rpmdp}
p(\rho_n)\propto\prod_{j=1}^k\alpha(n_j-1)!
\end{equation} where $n_1,\ldots,n_k$ is the cardinality of each cluster \cite{Lo1984}. Given the partition, in model (\ref{eq:dpm}) the observations assigned to different clusters are modelled as independent.

\section{Extension of the DP prior in regression settings}\label{litrev2} 
Many extension of the conventional DPM have been proposed in the literature. We describe here only the two relevant for our application. In particular, we work in a linear regression framework considering a vector of observations $\bm{y}=(y_1, \ldots, y_n)$, for which we assume the model \begin{equation}\label{eq:regression}
y_i\mid \bm{x}_i, \bm{\beta}_i, \lambda_i\sim \mbox{Normal}(y_i \mid \bm{x}_i\bm{\beta}_i, \lambda_i),
\end{equation}
where $\bm{x}_i$ is the $i-$th row of the $(n\times D)$ covariates matrix $\bm{X}$, $\bm{\beta}_i$ is a vector of subject-specific regression coefficients and $\lambda_i$ is the individual-level precision. 

The coefficients $\bm{\beta}_i$ are usually modelled using a multivariate Normal prior; however, while it is important to maintain a simple interpretation and guarantee easy computation for the posteriors, we would like to accommodate heterogeneity in the population and to allow for outliers, clustering and over-dispersion. This higher level of flexibility is often difficult to achieve using a single parametric random effect distribution. In addition, we aim at identifying the explanatory variables that influence the outcome the most. Thus, we extend the basic Bayesian regression framework and set a non-parametric prior instead.

\subsection{Covariate dependent clustering}
Recent developments in the RPM literature (particularly within a regression framework) have focussed on creating partitions of the observations that successfully take into account possible patterns within profiles determined by the combination of covariate values. This is particularly relevant for applications involving a large number of covariates that are expected to contain useful but correlated information about clustering and when the focus is on prediction. This type of models has been termed \textit{random partition model with covariates} (RPMx) \cite{mulleretal2010}.

In order to allow the partition of the observations to depend on the values of the covariates $\bm{x}$, several authors have proposed modifications of the basic clustering structure of the DP prior described in (\ref{eq:rpmdp}). For example, \cite{muelleretal1996} propose to model the response and covariate(s) jointly using a DPM model \cite{antoniak1974} and then use  the posterior conditional distribution of the response given the covariates and the partition of the observations to determine the relationship between $\bm{y}$ and $\bm{x}$. An alternative model is the so-called \textit{product partition model with covariates} (PPMx) \cite{mulleretal2011}. 
In this case a prior is specified directly on the partition:
$$p(\rho_n)\propto\prod_{j=1}^kc(S_j)$$
where $c(\cdot)$ is a ``cohesion'' function and $S_j$ is the set containing the observation labels belonging to cluster $j$ --- note that the DPM is a particular case, induced by choosing $c(S_j) \propto\alpha(n_j-1)!$. The PPMx extends the cohesion function to include a measure of similarity between covariates from different observations so that individuals with ``similar'' profiles are more likely to be grouped in the same cluster. 

Other extensions of DPM models which allow for covariate dependent clustering  are given by the DP with Generalized Linear Model (DP-GLM) \cite{hannahetal2011}, which accommodates different types of responses in a GLM framework; the Generalized Product Partition Model (GPPM) \cite{parketal2010}, which first clusters the individuals on the basis of the covariates and then uses the posterior probability of the partitions obtained as prior for the partitions built on the response; and the Profile Regression (PR) \cite{molitoretal2010}, in which a DPM model is specified for the covariates, which are then linked to the response through cluster-specific regression models.

\subsection{Variable (model) selection}
The DP framework can be also used to perform variable selection in the linear model (\ref{eq:regression}).  If we assume a DP process prior for the regression coefficients $\bm\beta_i$, then we obtain an infinite mixture of linear regression models. This implies a partition of the $n$ observations in clusters, each of which is characterised by specific values of the regression coefficients (\ie in the $j$--th cluster, $\bm\beta_j^*$ of length equal to the number of covariates, $D$). Thus, if we allow the parameter $\bm\beta_j^*$ to have some component(s) equal to zero with positive probability, we are effectively allowing for the possibility that some of the observed covariates in cluster $j$, denoted $\bm{x}_j^*$, do not affect the outcome $\bm{y}$ for the observations in that cluster, effectively performing  variable selection. 

For example, \cite{kimetal2009} propose a Bayesian non-parametric regression model employing a DP prior on the regression coefficients, but choosing as base measure a spike and slab distribution, to allow some coefficients of the regression to be zero (see \cite{georgeetal1993} and \cite{malsineretal2011} for a review on spike and slab distribution for variable selection). Similar approaches can be found in \cite{caietal2007} for linear mixed models and in \cite{dunsonetal2008} for binary regression. These methods perform covariate selection within each cluster and, as a consequence, observations are grouped on the basis of the effects of the covariates on the response. This strategy can lead to poor clustering in the case of a high-dimensional covariate space, and also to reduced predictive performance. This is because possible patterns within the covariates are not taken into account when clustering individuals.

Variable selection techniques have been developed also for the PPMx model \cite{mulleretal2011} and for the PR model \cite{papathomasetal2012}. In this setup, it is possible to perform simultaneously covariate dependent clustering and  selection of covariates that discriminate the most between clusters; however, no variable selection is performed to identify the covariates that mostly associated with the response of interest. 

Other examples of joint clustering and variable selection  can be found in \cite{tadesseetal2005, bruscoetal2001, mclachlanetal2002, fowlkesetal1988} among the others, but once again these methods  aim to highlight covariates that most explain the clustering structure, but not the relationship between covariates and outcome.

The main goal of this paper is to combine covariate dependent clustering and variable selection methods able to identify covariates that best explain the outcome variable, by generalising the approach in \cite{kimetal2009}. We refer to the proposed model as   Random Partition Model with covariate Selection (RPMS).

\section{\textbf{Random Partition Model with Covariate Selection (RPMS)}}\label{sec:srpmx}
In this section we develop the RPMS model  and briefly explain the Markov Chain Monte Carlo (MCMC) algorithm employed to perform posterior inference.

\subsection{Regression Model}
As discussed at the beginning of Section \ref{litrev2}, we use a linear regression model to explain the relationship between the response and the covariates. Let $\bm{y}=(y_1, \ldots, y_n)$ denote the response variable. Then, we assume
\begin{equation*}\label{eq:regression2}
y_i\mid \bm{x}_i, \bm{\beta}_i, \lambda_i\sim \mbox{Normal}(y_i \mid \bm{x}_i\bm{\beta}_i, \lambda_i).
\end{equation*}

Here, we assume that $x_{id} \in \{0,1\}$ for every $i=1,\ldots,n$ and  $d=1,\ldots,D$. Thus, we can interpret $\bm{X}$ as the matrix containing the information about the presence of $D$ symptoms for each of the $n$ patients; these symptoms are assumed to \textit{possibly} have an effect on the response $\bm y$. We focus on binary covariates, because in clinical settings symptoms are often recorded as binary indicators (in fact, that is the case in our motivating example). Extensions to other type of covariates is however trivial. 

The goal is to specify a prior structure that allows detecting a possible clustering structure based on symptoms profiles and then identifying which variables most influence (globally or in some clusters)  the response variable. 

\subsection{Model on the Covariates and Prior Specification }
To allow for covariate dependent clustering, we exploit ideas in \cite{muelleretal1996} assuming a probability model for the vectors of covariates:
\begin{equation*}\label{eq:priorx}
\bm{x}_i\mid \zeta_1,\ldots,\zeta_D \sim \prod_{d=1}^D\text{Bernoulli}(x_{id}\mid  \zeta_{id}).
\end{equation*}

In addition, we specify a joint DP prior distribution on $\bm{\zeta}_i=(\zeta_{i1},\ldots,\zeta_{iD})$ and $\bm{\beta}_i=(\beta_{i1},\ldots,\beta_{iD})$ :
\begin{eqnarray}
(\bm{\beta}_1,\bm{\zeta}_1), \ldots , (\bm{\beta}_n,\bm{\zeta}_n)\mid \mathcal{G} & \sim & \mathcal{G}\label{eq:priordp} \\
\mathcal{G} & \sim & \text{DP}(\alpha, \mathcal{G}_0) \nonumber
\end{eqnarray}
where $\alpha$ is the precision parameter and $\mathcal{G}_0$ is the base measure of the process. Recalling Section \ref{sec:intro}, the DP in (\ref{eq:priordp}) assigns a positive probability for two observations $i$ and $i'$ to have the same values $(\bm{\beta}_{i},\bm{\zeta}_i)=(\bm{\beta}_{i'},\bm{\zeta}_{i'})$. 
We denote with 
 $(\bm{\beta}^*,\bm{\zeta}^*)=((\bm{\beta}_1^*,\bm{\zeta}_1^*),\ldots,(\bm{\beta}_k^*,\bm{\zeta}_k^*))$  the unique values for the parameters. This construction implies that observations are clustered on the basis of both their covariates profile and the relationship between covariates and responses.

The model is completed by specifying a conjugate Gamma prior on the regression precision assuming $\lambda_1=\ldots=\lambda_n=\lambda$ and modelling $\lambda \sim \text{Gamma}(\lambda\mid a_{\lambda}, b_{\lambda})$, as well as using a Gamma hyper-prior on the concentration parameter of the DP \cite{escobaretal1995}: $\alpha \sim \text{Gamma}(a_{\alpha} , b_{\alpha})$. These are common prior choices as they enable easier computations.

\subsubsection{The Spike and Slab Base Measure}
The choice of the base measure of the DP is crucial. We assume that $\bm{\beta}_i$ and the $\bm{\zeta}_i$ are independent in the base measures. We choose a spike and slab distribution as base measure for the regression coefficients to perform  variable selection. Thus we define:

\begin{equation*}\label{eq:g0}
\mathcal{G}_0=\prod_{d=1}^D\{\left[\omega_d\delta_0(\beta_{\cdot d})+(1-\omega_d)\mbox{ Normal}(\beta_{\cdot d}\mid m_d, \tau_d)\right]\text{Beta}(\zeta_{\cdot d}\mid a_{\zeta} , b_{\zeta} )\},
\end{equation*}
which is simply the product measure on the space of the regression coefficients and of  the parameters defining the distribution of  the covariates. The notation $\beta_{\cdot d}$ and $\zeta_{\cdot d}$ highlights the fact that the base measure is assumed to be the same across the observations.  In $\mathcal{G}_0$, the first part within the square brackets is the spike and slab distribution, while $\delta_0(\beta_{\cdot d})$ is a Dirac measure that assigns probability 1 to the value 0. Thus a spike and slab distribution is a mixture of a point mass at 0 (in correspondence of which, $\beta_{\cdot d}=0$) and a Normal distribution, with weights given by $\omega_d$ and $(1-\omega_d)$, respectively. 

A conjugate base measure is employed for the covariate specific parameters $\zeta_{\cdot d}$ for ease of computations.
We assume the same hyperpriors for the parameters in $\mathcal{G}_0$ as in \cite{kimetal2009}. In particular we set  a spike and slab hyperpriors for each $\omega_d$:
\begin{eqnarray*}
\omega_1,\ldots,\omega_D\mid \pi_1,\ldots,\pi_D & \sim & \prod_{d=1}^D\{(1-\pi_d)\delta_0(\omega_d) + \pi_d\text{ Beta}(\omega_d\mid a_{\omega} , b_{\omega})\} \\
\pi_1,\ldots,\pi_D \mid a_{\pi},b_{\pi} & \sim & \prod_{d=1}^D\text{Beta}(\pi_d\mid  a_{\pi} , b_{\pi})
\end{eqnarray*}
The latter solution has been proposed by \cite{lucasetal2006} as the possibility to induce extra sparsity on the regression coefficients, encouraging those associated with the covariates having no effect on the response variable to shrink toward zero. As shown in \cite{kimetal2009}, it is possible to integrate out $\omega_d$ from  the base measure, obtaining:

\begin{displaymath}
\mathcal{G}_0=\prod_{d=1}^D\{[\pi_dw_{\omega}\delta_0(\beta_{\cdot d})+(1-\pi_dw_{\omega})\mbox{Normal}(\beta_{\cdot d}\mid m_d, \tau_d)]\text{Beta}(\zeta_{\cdot d}\mid a_{\pi} , b_{\pi})\}
\end{displaymath}
where $w_{\omega}=\frac{a_{\omega}}{(a_{\omega} + b_{\omega})}$.

We set $m_1=\ldots=m_D=0$; in addition, we use a a Gamma prior for the precision parameters of the Normal component of the spike and slab prior:
\begin{displaymath}
\tau_1,\ldots,\tau_D \mid a_{\tau},b_{\tau} \sim \prod_{d=1}^D\text{Gamma}(\tau_d\mid a_{\tau} , b_{\tau}).
\end{displaymath}
\subsection{{Posterior Inference}}\label{sec:pi}

Markov Chain Monte Carlo (MCMC) 	\cite{gilksetal1996} algorithms have been adopted to sample from the posterior distributions of interest. Since our model can be rewritten using a DPM formulation on the response and the covariates jointly, efficient Gibbs sampler schemes are available. We follow the auxiliary parameter algorithm proposed in \cite{neal2000}. This procedure updates first the vector of cluster allocations $\bm{s}$ and then separately all the cluster specific parameters and the parameters that do not depend on the cluster allocation. A detailed description of the updating steps  is reported in Appendix \ref{app:comp}. We present below a summary of the updating steps: 
\renewcommand{\labelenumi}{(\roman{enumi})} 
\begin{enumerate}
\item Update the membership indicator $\bm{s}=(s_1,\ldots,s_n)$ using the Gibbs sampling procedure for non-conjugate base measure by the auxiliary variable algorithm presented in \cite{neal2000}.
\item Update the precision of the DP, $\alpha$, exploiting the method implemented in \cite{escobaretal1995}, setting $\alpha \sim \text{Gamma}(\alpha \mid  a_{\alpha} , b_{\alpha})$ a priori.
\item  Update $\bm{\zeta}^*=(\bm{\zeta}_1^*,\ldots,\bm{\zeta}_k^*)$ from the full conditional distribution, given the new configuration of $\bm{s}$ in (i).
\item Update $\bm{\beta}^*=(\bm{\beta}_1^*,\ldots,
\bm{\beta}_k^*)$ from the full conditional posterior distribution, given the new configuration of $\bm{s}$ in (i).
\item Update $\bm{\pi}=(\pi_1,\ldots, \pi_D)$ from the full conditional distribution. To draw from this distribution  we implement the algorithm described in \cite{kimetal2009}. 
\item Update $\bm{\tau}=(\tau_1,\ldots,\tau_D)$ from the full conditional distribution.  
\item Update the precision of the regression $\lambda$ from the full conditional distribution. 
\end{enumerate}

\section{\textbf{Summarizing Posterior Output}}\label{sec:varsel}
The choice of a spike and slab base measure implies that the coefficients $\beta^*_{jd}$ have positive probability to be equal to zero. We propose here two ways of summarising the MCMC output  that highlight the effect of using a spike and slab prior distribution. These two methods are then applied to the real data example in the following section.

In our framework a covariate can be explanatory for a cluster and not for another. Thus, a first method to analyse the results would be to compute the probability that the $d-$th covariate has explanatory power in cluster $j$, \ie $p(\beta^*_{jd}\neq 0)$, given a partition of the observations in clusters. 
The literature proposes a variety of methods for extracting a meaningful partition from the MCMC output \cite{dahl2009,molitoretal2010}. In our application we have decided to report the partition obtained by minimizing the Binder loss function \cite{binder1978}. 
Then, conditioning on the selected partition, we can compute the posterior distribution of the regression coefficients for each cluster, together with the probability of inclusion of a certain covariate, \ie $1-p(\beta_{jd}^*=0\mid \hat{\bm{s}}, \cdot)$, where $\hat{\bm{s}}$ is the Binder configuration.  

A second way of summarising the posterior output from a variable selection perspective is based on predictive inference. Let us consider the situation in which a new patient enters the study with profile $\tilde{\bm{x}}$. Using the proposed approach, the posterior distribution of the regression coefficients depends on the structure of the patient's profile. This is due to the fact that the cluster allocation depends on it. In fact, in RPMS the predictive distribution of the cluster allocation is:
\begin{equation}\label{eq:preddp}
p(\tilde{s}\mid \tilde{\bm{x}}, \ldots) \propto \left\{\begin{array}{ll} 
\displaystyle n_j \prod_{d=1}^Dg_{jd}(\tilde{x}_d) & \textrm{for $j=1,\ldots, k$} \\
\displaystyle \alpha {\prod_{d=1}^Dg_{0d}(\tilde{x}_d)} & \textrm{ for $j=k+1$}
\end{array} \right.
\end{equation}
where $\tilde{\bm{x}}=(\tilde{x}_1,\ldots,\tilde{x}_{D})$ and $\tilde{s}$ are the profile for the new patient and its cluster allocation, respectively. In addition, $g_{jd}(\tilde{x}_d)=\zeta_{jd}^{*\tilde{x}_d}(1-\zeta_{jd}^*)^{(1-\tilde{x}_d)}$ and $g_{0d}(\tilde{x}_d)=(\int_0^{1}q^{(\tilde{x}_d+a_{\zeta}-1)}(1-q)^{(b_{\zeta}-\tilde{x}_d)}\text{d}q)/(\int_0^{1}u^{(a_{\zeta}-1)}(1-u)^{b_{\zeta}-1}\text{d}u)$ are the likelihood for the new observation to belong in cluster $j$ and the prior predictive distribution of the new observation, respectively. The probability in (\ref{eq:preddp}) comes directly from the predictive scheme of the DP \cite{blackwell1973}.

Hence, we focus on $p(\tilde{\beta}_{d}=0\mid \tilde{\bm{x}}, \tilde{\bm{s}}, \cdot)$. This probability can be approximated using the MCMC samples. Moreover, it is possible to look at the predictive distribution of the response, namely $\tilde{y}$,  that by construction depends on the variable selection.
Notice that the model in \cite{kimetal2009} does not specify a model on the covariates, and consequently clustering does not depend on the covariates profiles. 

Alternatively, we could compute the posterior probability of $p(\beta_{1d}^*=\ldots=\beta_{kd}^*=0\mid \cdot)$ to summarize the overall importance of the $d-$th covariate. This posterior probability can be approximated empirically by calculating the proportion of iterations in the MCMC run in which the regression coefficient for the  $d-$th covariate is equal to zero in all the clusters: $\beta_{1d}^*=\ldots=\beta_{kd}^*=0$. 

\section{\textbf{Lower Urinary Tract Symptoms (LUTS) data}}\label{sec:luts}
In this section we present the results of the application of the RPMS to the LUTS database. First, we briefly describe the database, then we give details of the choice of the hyper-parameters and MCMC settings. We briefly introduce the competitor model and finally we report the results for clustering and variable selection.

\subsection{Data}\label{sec:database}
We consider data on 1341 patients extracted from the LUTS database collected at the LUTS clinic, Whittington Hospital Campus, University College London. For each patient, the presence of 34 LUTS has been recorded together with the White Blood Cell count (WBC) in a sample of urine. The patients are women aged over 18, affected by LUTS. We consider data at the first attendance visit. 

It is of clinical interest to investigate the relationship between LUTS and Urinary Tract Infection (UTI), where the latter is measured by the number of WBC. In particular, a value of $\text{WBC}\geq1$ is indicative of the presence of infection. 

The symptoms are stored as binary variables (1 indicates the presence of the symptoms and 0 the absence). We report  the frequency distribution of the symptoms in the 1341 patients in Table \ref{tab:symp}. The symptoms can be grouped into four categories: urgency symptoms (symptoms from 1 to 8), stress incontinence symptoms (9 to 14), voiding symptoms (15 to 21) and pain symptoms (22 to 34). 

\begin{table}[!h]\label{tab:symp}
\caption{Lists of the 34 symptoms with the frequency of occurrence.}
\fontsize{10}{12}\selectfont
\centering
\begin{tabular}{lclc}
\textbf{Symptom}           & \textbf{Frequency} & \textbf{Symptom}          & \textbf{Frequency} \\ 
1) Urgency incontinence  & 0.4146             & 18) Straining to void         & 0.0828             \\
2) Latchkey urgency           & 0.4280              & 19) Terminal dribbling        & 0.1641             \\
3) Latchkey incontinence      & 0.2304             & 20) Post void dribbling & 0.0820              \\
4) Waking urgency      & 0.5496             & 21) Double voiding            & 0.1193             \\
5) Waking incontinence & 0.2595             & 22) Suprapubic pain           & 0.1611             \\
6) Running water urgency      & 0.2901             & 23) Filling bladder pain   & 0.2148             \\
7) Running water incontinence & 0.1365             & 24) Voiding bladder pain      & 0.0567             \\
8) Premenstrual aggravation  & 0.0515             & 25) Post void bladder pain     & 0.0723             \\
9) Exercise incontinence      & 0.1462             & 26) Pain fully relieved by voiding   & 0.0634             \\
10) Laughing incontinence      & 0.1536             & 27) Pain partially relieved by voiding  & 0.1260              \\
11) Passive incontinence       & 0.0783             & 28) Pain unrelieved by voiding     & 0.0164             \\
12) Positional incontinence    & 0.0850              & 29) Loin pain                 & 0.2081             \\
13) Standing incontinence      & 0.0895             & 30) Iliac fossa pain                   & 0.0895             \\
14) Lifting incontinence       & 0.1104             & 31) Pain radiating to genitals              & 0.0865             \\
15) Hesitancy                  & 0.1797             & 32) Pain radiating to legs                  & 0.0649             \\
16) Reduced stream             & 0.1909             & 33) Dysuria                   & 0.1484             \\
17) Intermittent stream        & 0.1514             & 34) Urethral pain             & 0.0507  \\        
\end{tabular}
\end{table} 

In this paper we focus only focus on patients with UTI (WBC$\geq 1$). We consider a logarithmic transformation of the WBC data and model the log-transformed  WBC using a Normal distribution. The left panel in Figure \ref{fig:brier} displays the kernel density estimation of the log-transformed WBC.

\subsection{Prior Specification}
The hyperparameters of the spike and slab prior in the base measure are set as follows: $a_{\omega}=a_{\pi}=1$, $b_{\omega}=b_{\pi}=0.15$, $a_{\tau}=b_{\tau}=1$ and $a_{\zeta}=b_{\zeta}=1$. We note here that we set vague prior beliefs on the distribution of the parameters, except for the prior on $\pi_d$ and $\omega_d$ to make computations more stable. The hyperparameters $a_{\lambda},b_{\lambda}$ for the precision $\lambda$ in the regression density are  set both to be equal to 1. Finally, for the prior on concentration parameter of the DP we use $a_{\alpha}=b_{\alpha}=1$. 

We run the MCMC sampler for 10\,000 iterations with a burn-in period of 1000 iterations. Details on the algorithm are given in Appendix \ref{app:comp}. To update the membership indicator, we use the auxiliary variable approach described in \cite{neal2000} that requires the choice of a tuning parameter $M$. In our experience, $M=100$ gives a good trade-off between execution time and efficiency of the Gibbs sampler. The convergence of the chains is assessed by trace plots and by the Gelman and Rubin's convergence diagnostic \cite{gelmanetal1992}, the latter for the parameters that do not depend on the cluster assignment.

\subsection{The competitor model: SSP}
In order to highlight the potential and advantages of RPMS, we compare its results with the model described in \cite{kimetal2009}, which we believe is the closest competitor.  For simplicity, we refer to this model as SSP (Spike and Slab Prior). This assumes the same Normal specification for the WBC counts and a DP prior on the regression coefficients with a spike and slab base measure for the regression coefficients. 

The difference with our own specification consists in the fact the the SSP treats the covariates as given, instead of associated with a probability distribution. This implies that in the SSP the base measure of the DP prior reduces to:
\begin{displaymath}
\mathcal{G}_0=\prod_{d=1}^D\{\omega_d\delta_0(\beta_{\cdot d})+(1-\omega_d)N(\beta_{\cdot d}\mid m_d, \tau_d)\}.
\end{displaymath}

We follow the same strategy adopted for the RPMS of integrating out from each part of the base measure the $\omega_d$. The model described in \cite{kimetal2009} involves also the use of a DP prior on the precision in the regression model, but for a fair comparison with the RPMS we use a version of the model without this further complication. Moreover, the results obtained by the SSP with or without the DP prior distribution on $\lambda$ have not shown to be significantly different. In the MCMC algorithm, we use the same initial values, tuning parameters and number of iterations utilised for the RPMS to obtain the posterior distributions. 

\subsection{Clustering outputs}
The proposed method, as explained above, employs a DP prior for the regression coefficients and for the parameters governing the profile distribution. The main consequence is that the implied clustering is influenced by both the distribution of the $y=\log(\mbox{WBC})$ and by possible patterns within the profiles. 

\begin{figure}[!h]
\begin{center}
\includegraphics[width = 3in]{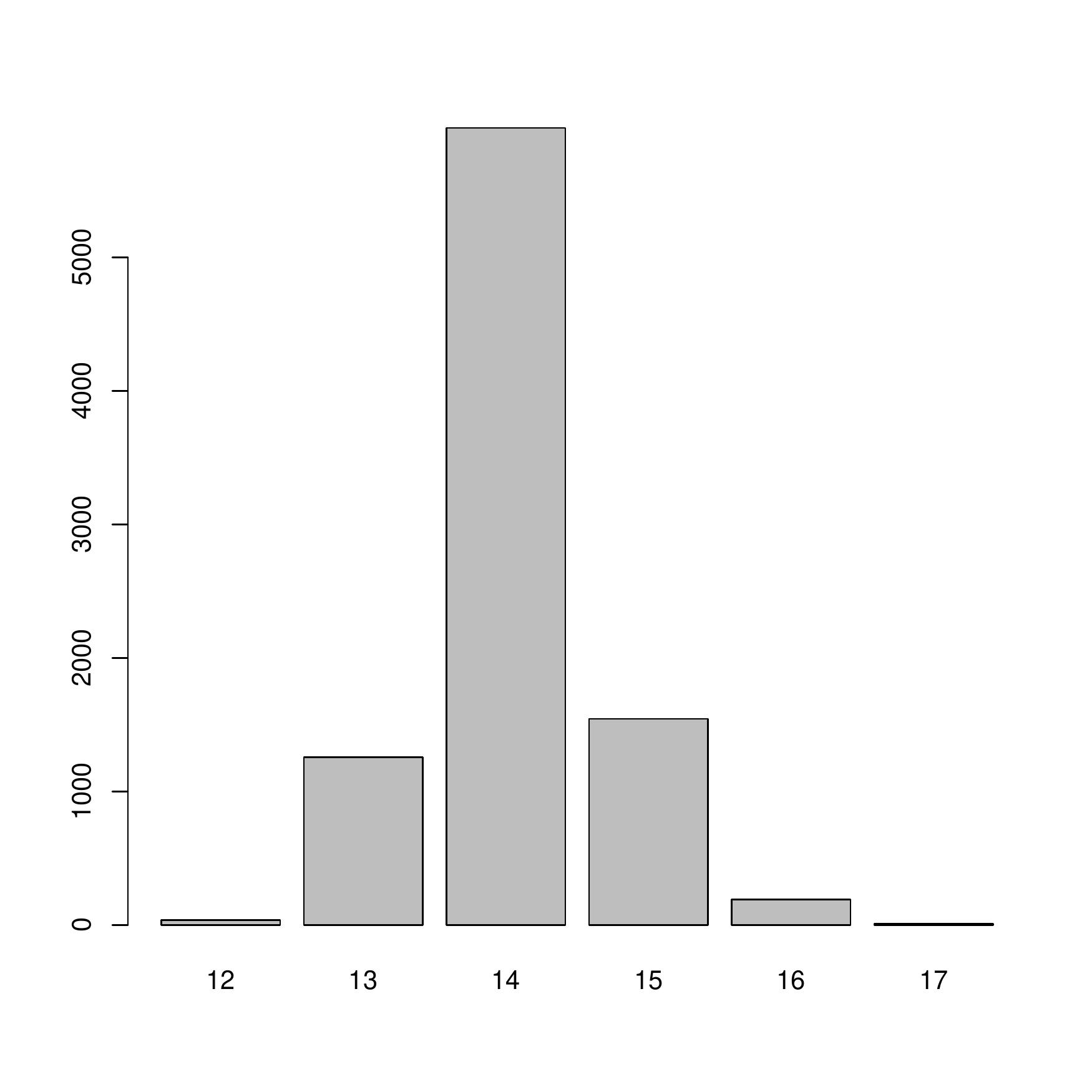}
\includegraphics[width = 3in]{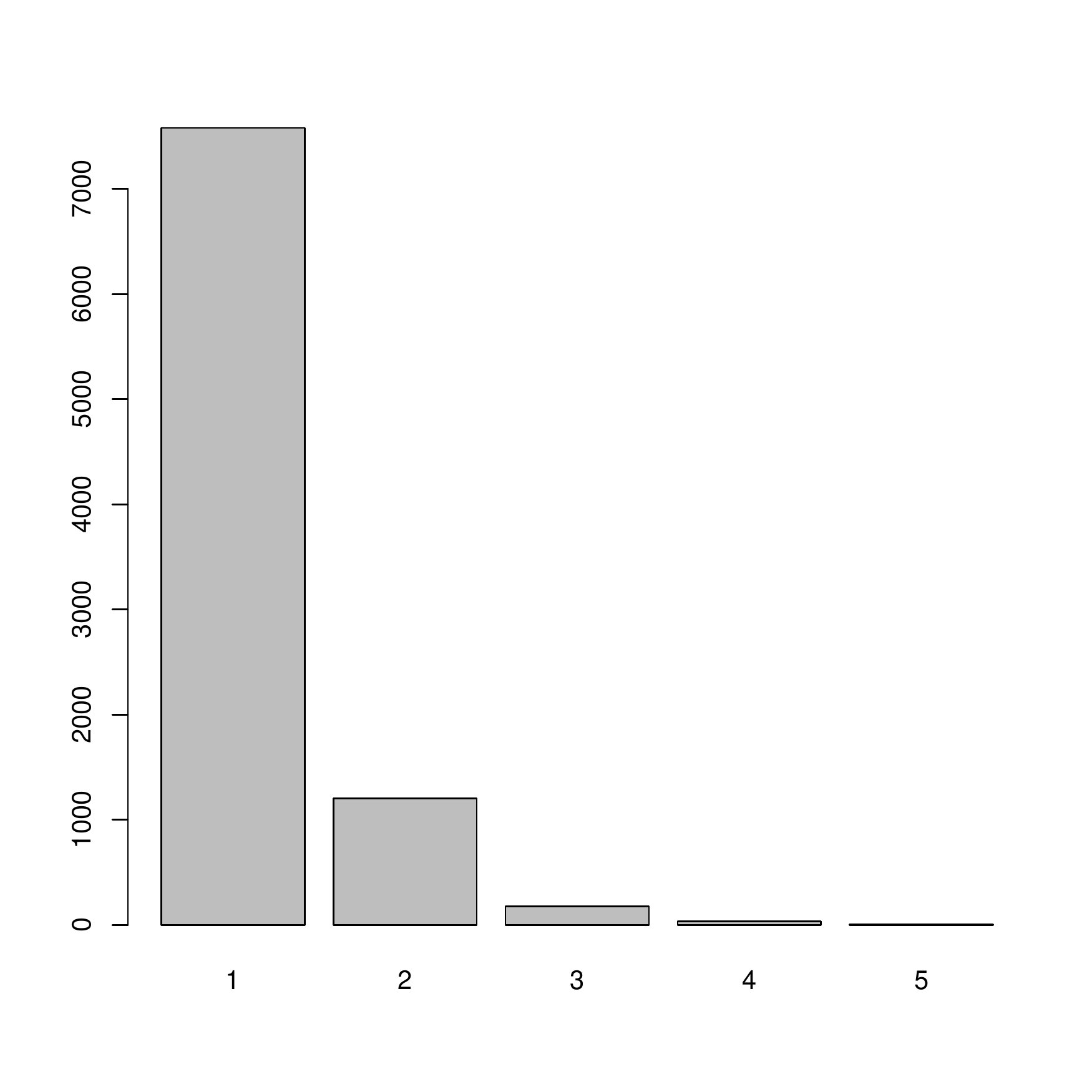} 
\put(-436,+85){\rotatebox{90}{\small{Frequency}}}
\put(-355, +9){\small{$k$ in RPMS}}
\put(-120, +9){\small{$k$ in SSP}}
\caption{Posterior distribution of the number of clusters $k$ for RPMS and for SSP models.}
\label{fig:cluster}
\end{center}
\end{figure}

Figure \ref{fig:cluster} reports the posterior distribution for $k$, \ie the number of clusters, from the RPMS model. The configuration involving 14 clusters is clearly the one with the highest probability. This is the first significant difference with the SSP model that has a clear mode at $k=1$. This is due to the fact that the SSP takes into account only the variability in the regression coefficients. In the RPMS the covariates contribute to inform  the partition of the observations. 

To summarise the posterior inference on the clustering we report the partition which minimises the Binder loss function~\cite{binder1978}
\begin{equation}\label{eq:binder}
L(\bm{\hat{s}}, \bm{\bar{s}})=\sum_{i<i^\prime}\left(\ell_1 I_{\{\hat{s}_i\neq \hat{s}_{i^\prime}\}}I_{\{\bar{s}_i = \bar{s}_{i^\prime}\}} + \ell_2 I_{\{\hat{s}_i = \hat{s}_{i^\prime}\}}I_{\{\bar{s}_i \neq \bar{s}_{i^\prime}\}}\right),
\end{equation}
where $\bm{\hat{s}}$ is a proposed partition, while $\bm{\bar{s}}$ indicates the true partition. The choice of the constants $\ell_1$ and $\ell_2$ allows to express the preference for a small number of large clusters or for a large number of small clusters, respectively. In our application we set $\ell_1=\ell_2=1$ penalising both terms equally. In this application the true partition is unknown, whereas  proposed partitions are represented by draws from the posterior distribution of the membership indicators.

The posterior expectation of (\ref{eq:binder}) is
\begin{displaymath}
\mbox{E}(L(\bm{\hat{s}}, \bm{\bar{s}})\mid \text{Data})=\sum_{i<i^\prime}\mid I_{\{\hat{s}_i=\hat{s}_{i^\prime}\}}-\gamma_{ii^\prime}\mid 
\end{displaymath}
where $\gamma_{ii^\prime}=\mbox{E}(I_{\{\bar{s}_i=\bar{s}_{i^\prime}\}}\mid \text{Data})$ and it can be consistently estimated computing the empirical probability of observations $i$ and $i^\prime$ to be clustered together across the observed MCMC iterations. Minimising the Binder loss function in our case leads to a configuration with 14 clusters, with the 9 largest clusters containing 92.5\% of the observations. 

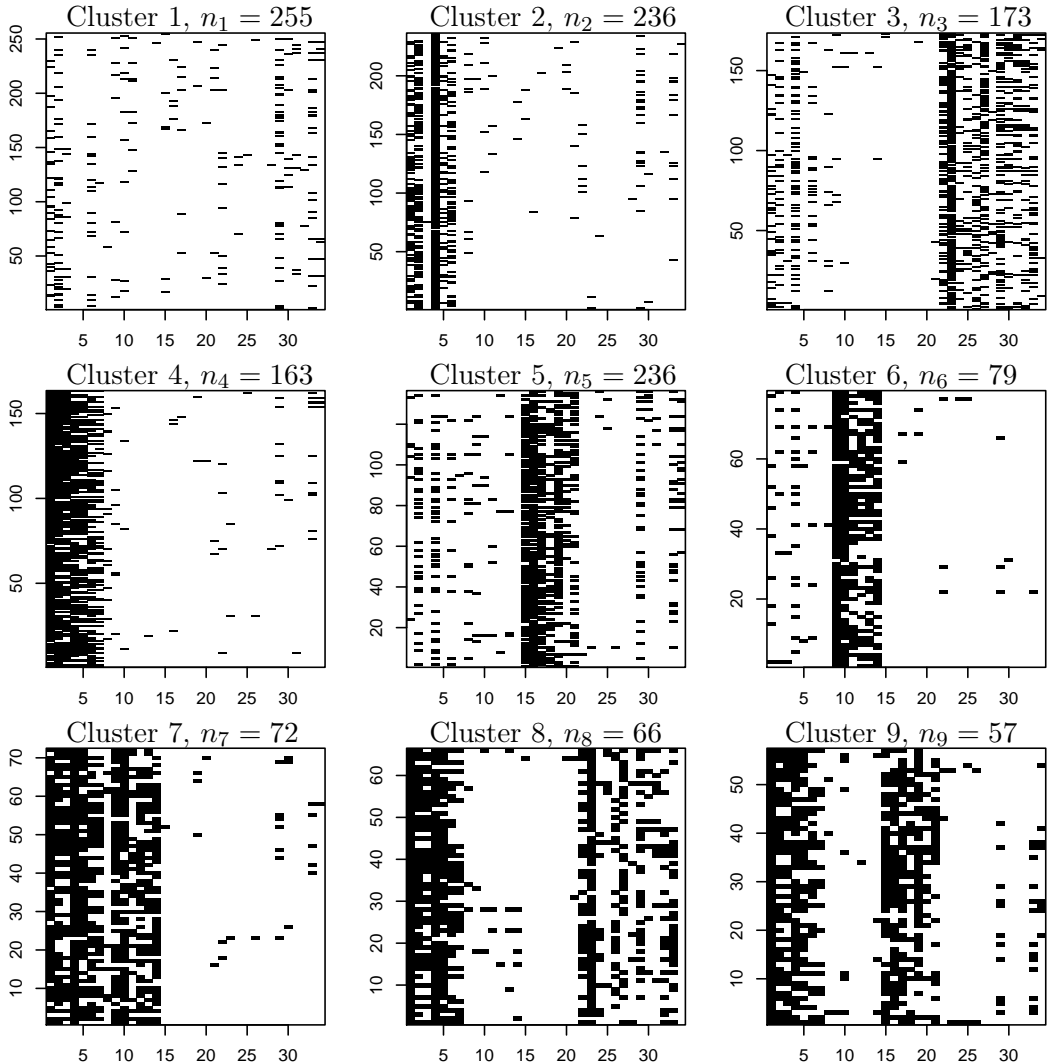
\begin{figure}[!h]
\begin{center}
\includegraphics[scale=0.8]{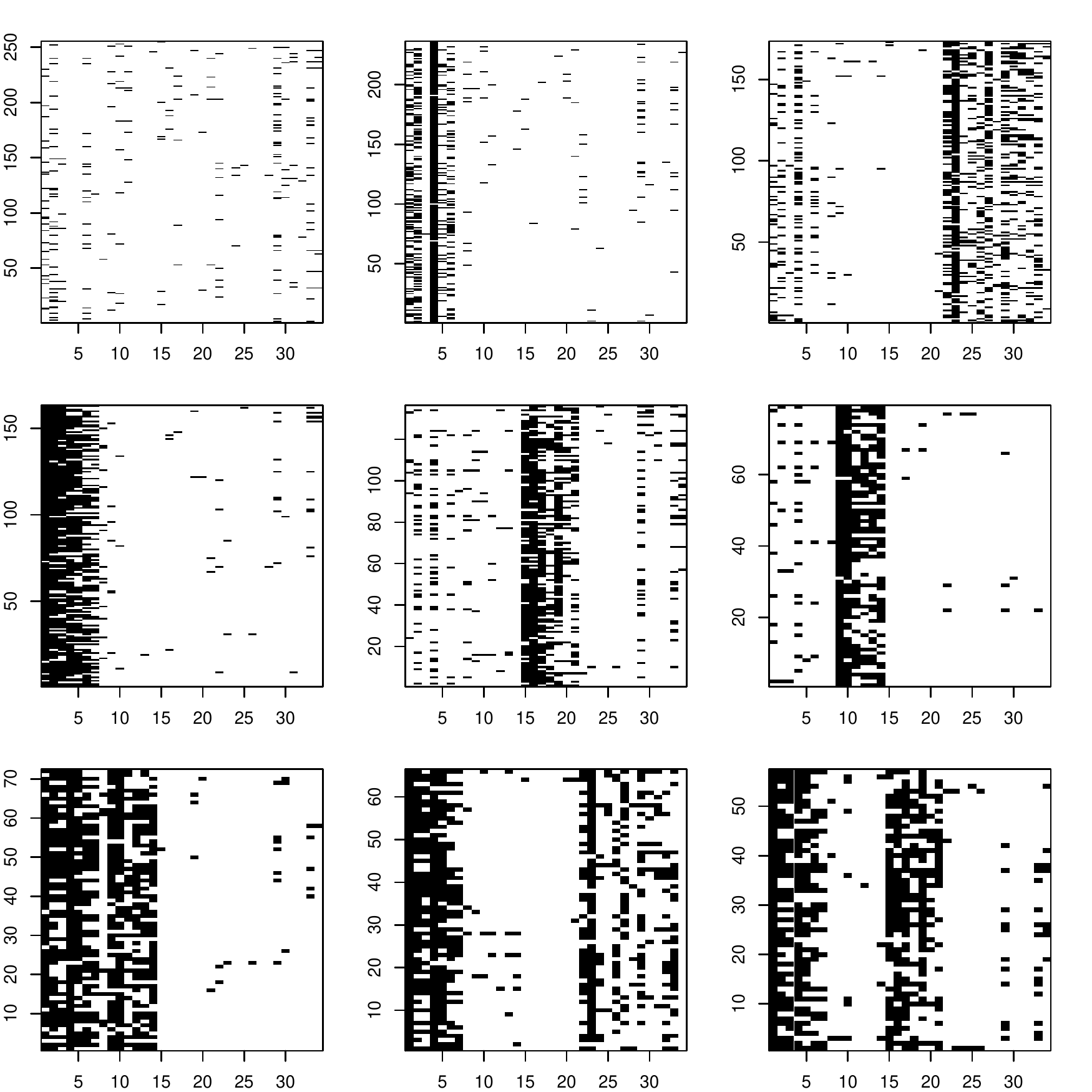}
\put(-382, +392){\small{Cluster 1, $n_1=255$}}
\put(-246, +392){\small{Cluster 2, $n_2=236$}}
\put(-113, +392){\small{Cluster 3, $n_3=173$}}
\put(-382, +257){\small{Cluster 4, $n_4=163$}}
\put(-246, +257){\small{Cluster 5, $n_5=236$}}
\put(-113, +257){\small{Cluster 6, $n_6=79$}}
\put(-382, +122){\small{Cluster 7, $n_7=72$}}
\put(-246, +122){\small{Cluster 8, $n_8=66$}}
\put(-113, +122){\small{Cluster 9, $n_9=57$}}
\caption{Symptom indicators (black) for the 9 biggest clusters of the partition obtained by minimizing the Binder loss function. The horizontal axis of each panel (corresponding to a cluster) displays the index of the symptoms, whereas the index of the patients in each cluster is on the vertical axis.  For each cluster the cardinality is also displayed.}
\label{fig:binder}
\end{center}
\end{figure}

Figure \ref{fig:binder} displays the presence of the 34 symptoms (on the columns) for the patients assigned to each of the nine largest clusters. Recalling that the symptoms can be grouped into four main categories, \ie urgency, stress incontinence, voiding and pain symptoms, we can see that the largest cluster contains patients with a small number of symptoms belonging to all the four categories, or with no symptoms at all. In the second largest cluster, almost all patients present the fourth class of symptoms and almost none the third one. A high frequency of the other urgency symptoms is also evident. The third largest cluster includes patients with a high frequency of pain symptoms together with urgency symptoms (even though with a lower probability). The fourth cluster is characterised by a high frequency of urgency symptoms; the fifth cluster by a high frequency of voiding symptoms; the sixth cluster by a high frequency of incontinence symptoms; the seventh cluster by a high frequency of urgency and incontinence symptoms; the eighth cluster by a high frequency of urgency and pain symptoms and the ninth cluster by a high frequency of urgency and voiding symptoms.   

This distribution of the symptoms across the Binder configuration suggests that the symptoms classes are informative for the partition.  Consequently, it is likely that each combination of symptoms  has a particular effect on the WBC counts distribution: this is because cluster specific regression coefficients are associated to cluster specific probabilities of having the symptoms.

\subsection{Variable selection outputs}
Our proposed model performs simultaneously clustering and variable selection. It is of clinical interest to check which symptoms have a significant impact on the response variable. In our case, this means checking which symptoms are more likely to be predictive of an underlying infection. 

In this section we will use the two ways of summarising the variable selection information produced by the RPMS that have been described in section \ref{sec:varsel}. The first one is based on the Binder estimate of the clustering configuration, while the second focusses on the predictive distribution for a new patient. We first report the posterior probability of each symptoms to be included in the model, conditional on the Binder estimate of the clustering configuration.

\begin{figure}[!h]
\begin{center}
\includegraphics[scale=0.9]{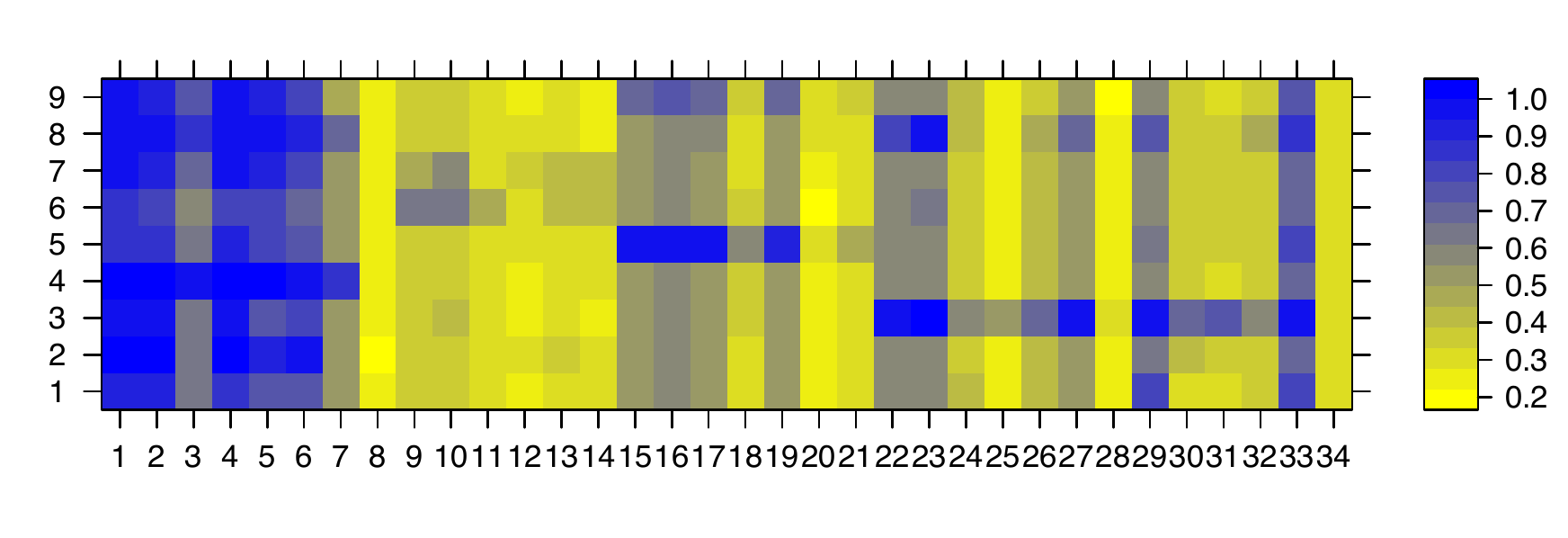}
\put(-450,+70){\rotatebox{90}{\small{Cluster}}}
\put(-250, +8){\small{Symptom}}
\caption{Probability of inclusion, \ie $\beta\neq0$, for each symptoms in the 9 biggest clusters of the partition estimated by minimizing the Binder loss function.}
\label{fig:inclusion}
\end{center}
\end{figure}

Figure \ref{fig:inclusion} displays the probability of inclusion, \ie $1-p(\beta_{jd}^*=0\mid \bm{\hat{s}}, \cdot)$ for the 9 largest clusters according to the Binder estimate ordered by size. For example, let us consider the fifth row (which refers to the fifth cluster). From Figure \ref{fig:binder}, we see that this cluster contains mainly the symptoms from 15 to 22 (cfr.\ the list in Table \ref{tab:symp}). Consequently, in Figure \ref{fig:inclusion} the probability that symptoms 15, 16, 17, 19 are included in the regression model is close to 1. On the contrary, for symptoms 18, 20, 21 and 22 the probability of being included is low. Figure \ref{fig:inclusion} also suggests the importance of the symptoms in the urgency class and of \textit{dysuria} and \textit{loin pain} within the pain class.

The second way of presenting variable selection results is by considering predictive inference. Recall that for a new patient entering the study the distribution of the regression coefficients depends on his profile, \ie $\tilde{\bm{x}}$, through the cluster assignment. This does not happen in the SSP, in which the predictive probability for the cluster assignment depends exclusively on the cardinality of the clusters.

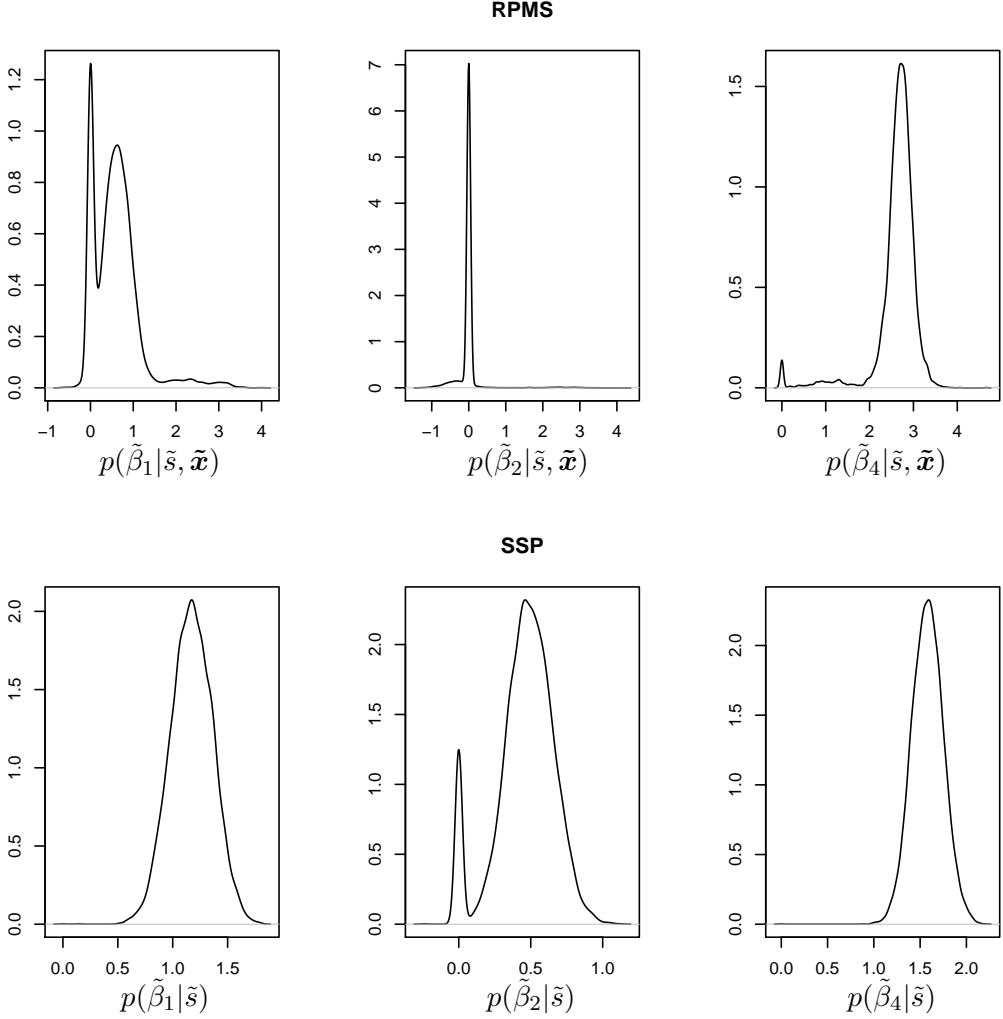
\begin{figure}[!h]
\begin{center}
\includegraphics[scale=0.8]{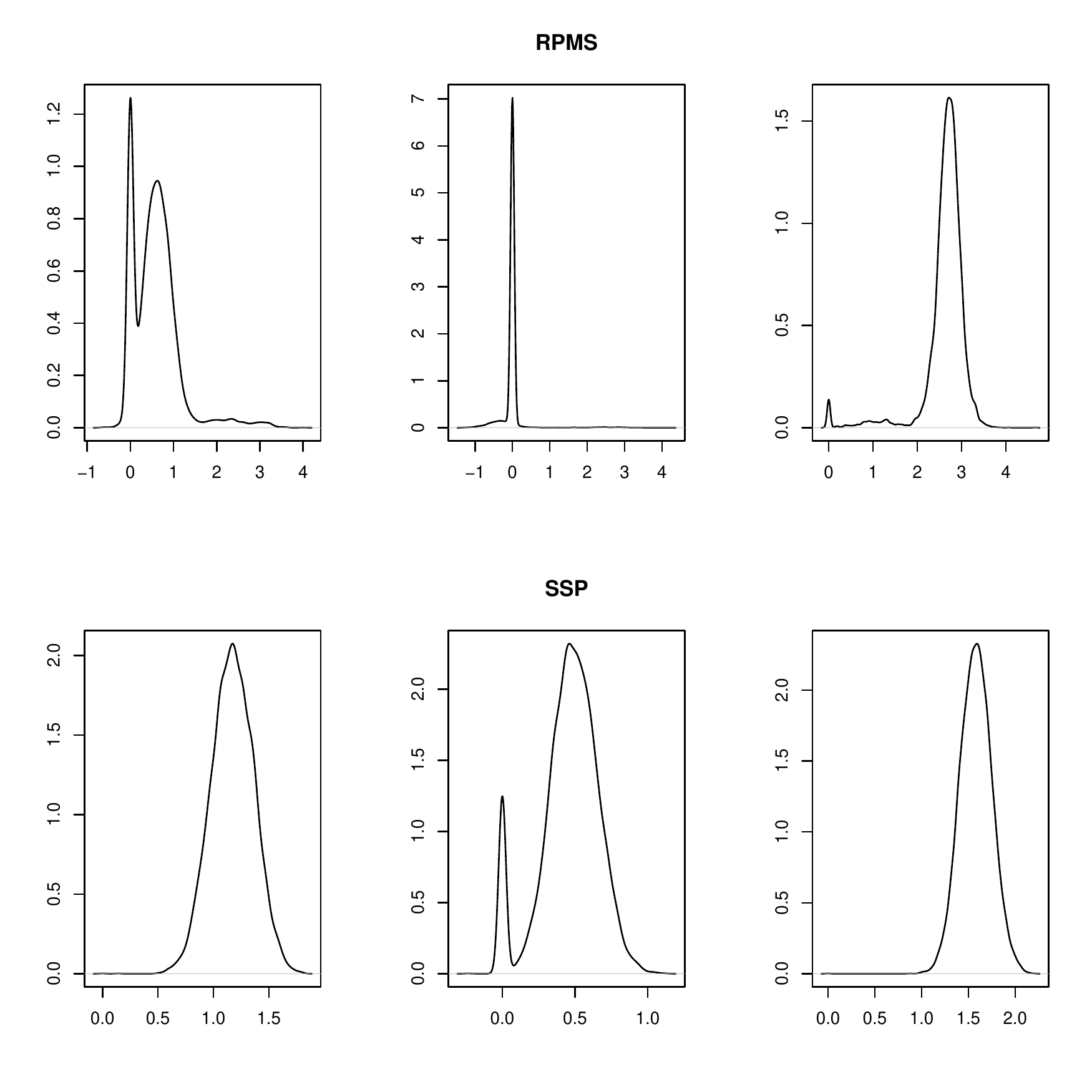}
\put(-345, +13){\small{$p(\tilde{\beta}_1|\tilde{s})$}}
\put(-208, +13){\small{$p(\tilde{\beta}_2|\tilde{s})$}}
\put(-73, +13){\small{$p(\tilde{\beta}_4|\tilde{s})$}}
\put(-353, +215){\small{$p(\tilde{\beta}_1|\tilde{s}, \bm{\tilde{x}})$}}
\put(-215, +215){\small{$p(\tilde{\beta}_2|\tilde{s}, \bm{\tilde{x}})$}}
\put(-81, +215){\small{$p(\tilde{\beta}_4|\tilde{s}, \bm{\tilde{x}})$}}
\caption{Kernel density estimation of the posterior distribution for $\tilde{\beta}_1$, $\tilde{\beta}_2$ and $\tilde{\beta}_4$ given the new profile $\bm{\tilde{x}}$ with $x_1=x_2=x_4=1$. The first row refers to the RPMS model, while the second row refers to the SSP model. For SSP the posterior distribution for the regression coefficients does not depend on $\bm{x}$.}
\label{fig:bepost}
\end{center}
\end{figure}

To illustrate the last considerations, we take $\tilde{\bm{x}}$ include the presence of symptoms 1, 2 and 4 from Table \ref{tab:symp}. Figure \ref{fig:bepost} shows the density estimation of the posterior distribution of the regression coefficients related to the three symptoms in $\bm{\tilde{x}}$. We present the output from both the  RPMS and the SSP. The evident differences between the distributions are due to the fact that in the SSP the regression coefficients do not depend on the profile, which they do in the RPMS. Consequently, in the SSP the posterior distribution of the regression coefficients is the same for every (new) patient, while in the RPMS it can vary, depending on the covariates profile. Moreover, in the RPMS  the spike and slab prior distribution can be seen as a within--cluster prior. 

\begin{figure}[!h]
\begin{center}
\includegraphics[width=3in]{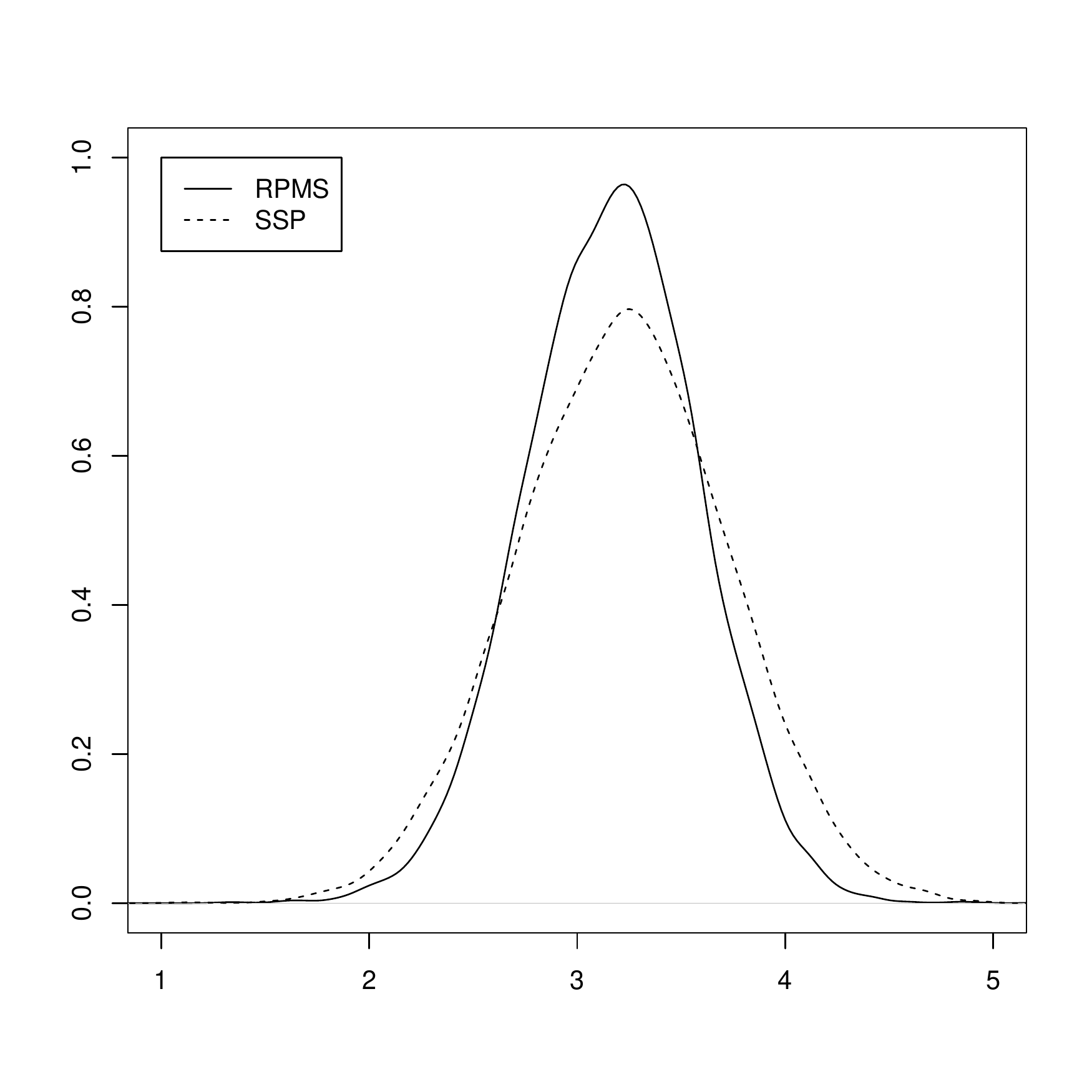}
\includegraphics[width=3in]{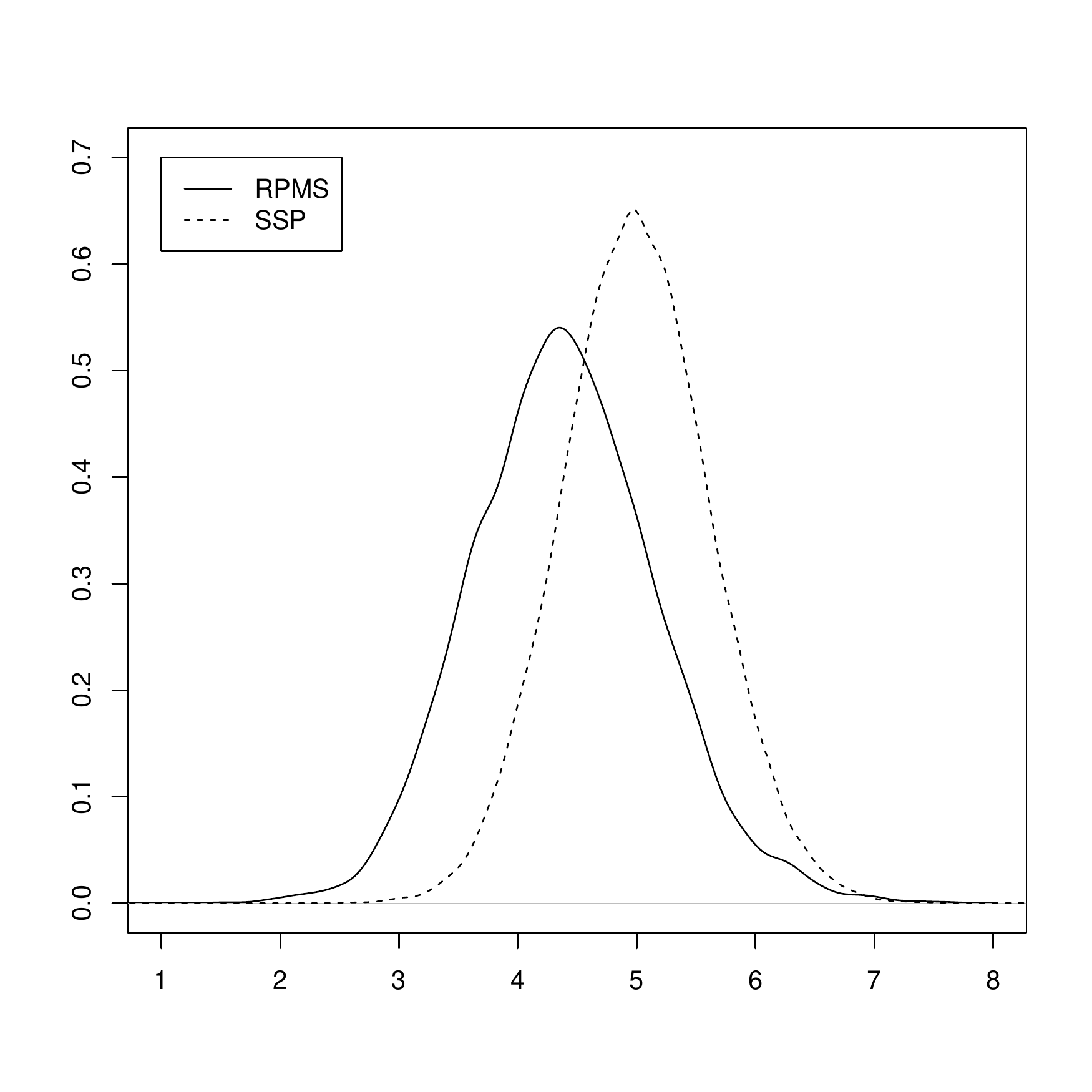}
\put(-340, +8){\small{$p(\tilde{y}|\bm{\tilde{x}})$}}
\put(-121, +8){\small{$p(\tilde{y'}|\bm{\tilde{x'}})$}}
\caption{Kernel density estimation of the predictive distribution of $\tilde{y}$ given profile $\bm{\tilde{x}}$ with $x_1=x_2=x_4=1$ and of $\tilde{y'}$ given  profile $\bm{\tilde{x'}}$ with $x_1=x_2=x_3=x_4=x_5=x_6=x_7=x_{22}=x_{23}=x_{27}=x_{28}=x_{32}=x_{33}=1$ }
\label{fig:ypred}
\end{center}
\end{figure}

The different posterior distributions of the regression coefficients for the SSP and the RPMS have obviously an impact on the predictive distribution of $\tilde{y}$. Figure \ref{fig:ypred} displays the predictive distribution of the response given a profile $\bm{\tilde{x}}$ (we assume these are the same as those considered in Figure \ref{fig:bepost}) and for a different profile $\bm{\tilde{x}'}$, which is characterized by a large number of symptoms: $x_1=x_2=x_3=x_4=x_5=x_6=x_7=x_{22}=x_{23}=x_{27}=x_{28}=x_{32}=x_{33}=1$. 

In the first case, the distributions obtained from the SSP and the RPMS have similar means, but the one estimated by the RPMS seems to have smaller variance. On the other hand for $\bm{\tilde{x}'}$  the predictive distributions seem more substantially different. 

In order to determine whether the proposed model leads to improved predictions we employ the Brier statistic \cite{brier1950}. This statistic assesses the quality of predictions when the response variable is binary:
\begin{equation*}\label{eq:brier}
\text{Brier} = \frac{1}{n}\sum_{i=1}^{n}(f_i-o_i)^2,
\end{equation*}
where $o_i$ is a binary observation and $f_i$ is its predicted probability. 

In our case the response variable is continuous, so in order to apply this statistic we discretize the response variable.  In this case we are interested in predicting whether the WBC count of an individual is above or below a specific threshold. We use the quartiles of the observed WBC counts as thresholds. Then, the predicted $f_i$ will be the probability of obtaining values larger than the specified threshold. For each patient $i$ in the sample, we estimate $f_i$ from the predictive distribution. Hence, the Brier statistic becomes:  
\begin{equation*}\label{eq:brier1}
\text{Brier}_{(q)} = \frac{1}{n}\sum_{i=1}^{n}\left(f^{(q)}_i-y^{(q)}_i\right)^2
\end{equation*}
where $y^{(q)}_i$ is the $i-$th response discretised with respect to the $q-$th quartile, whilst $f^{(q)}_i$ is the predictive probability that an individual has a WBC value larger than the $q-$th quartile. The sum is taken over all the patients in the sample. 

\begin{figure}[!h]
\begin{center}
\includegraphics[width=3in]{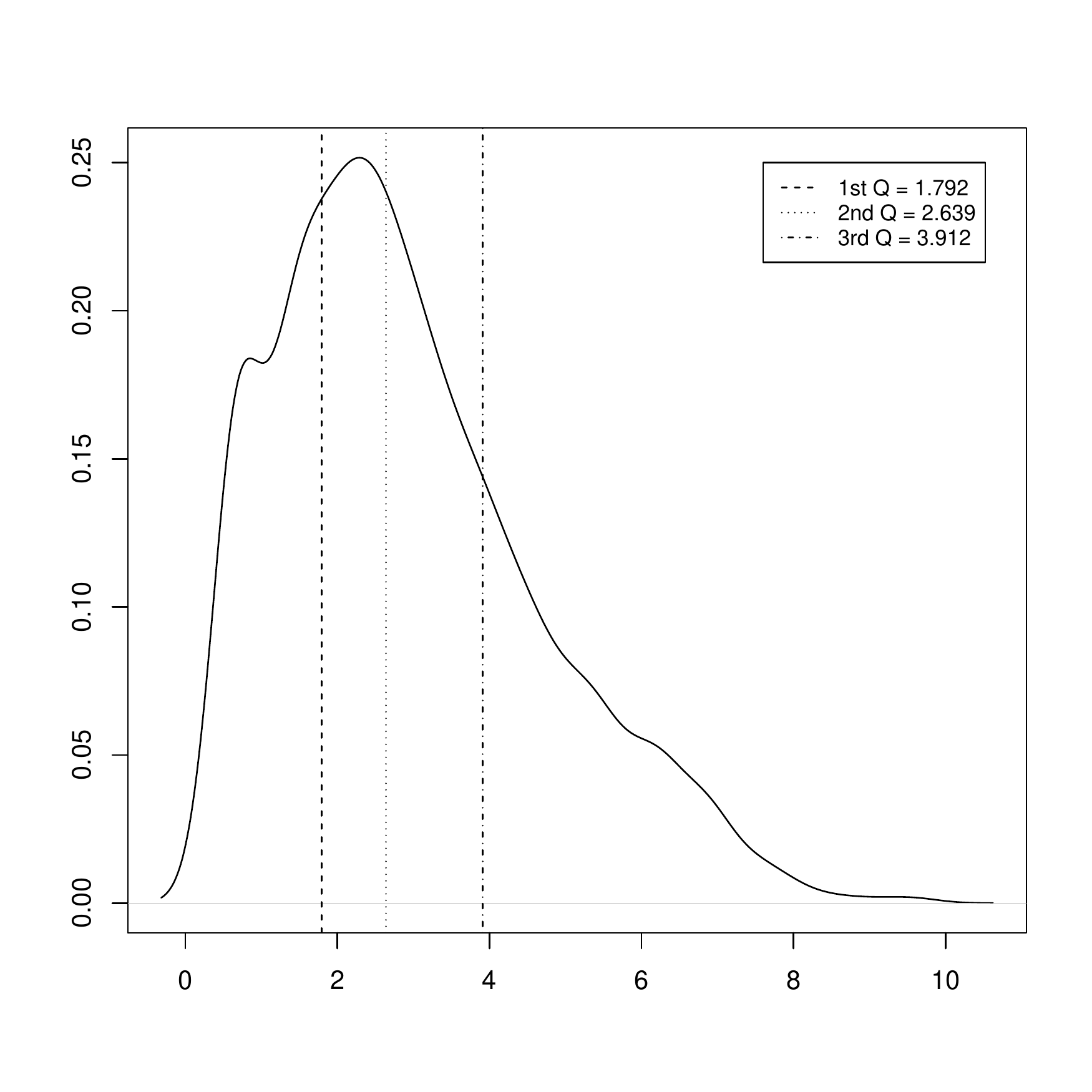}
\includegraphics[width=3in]{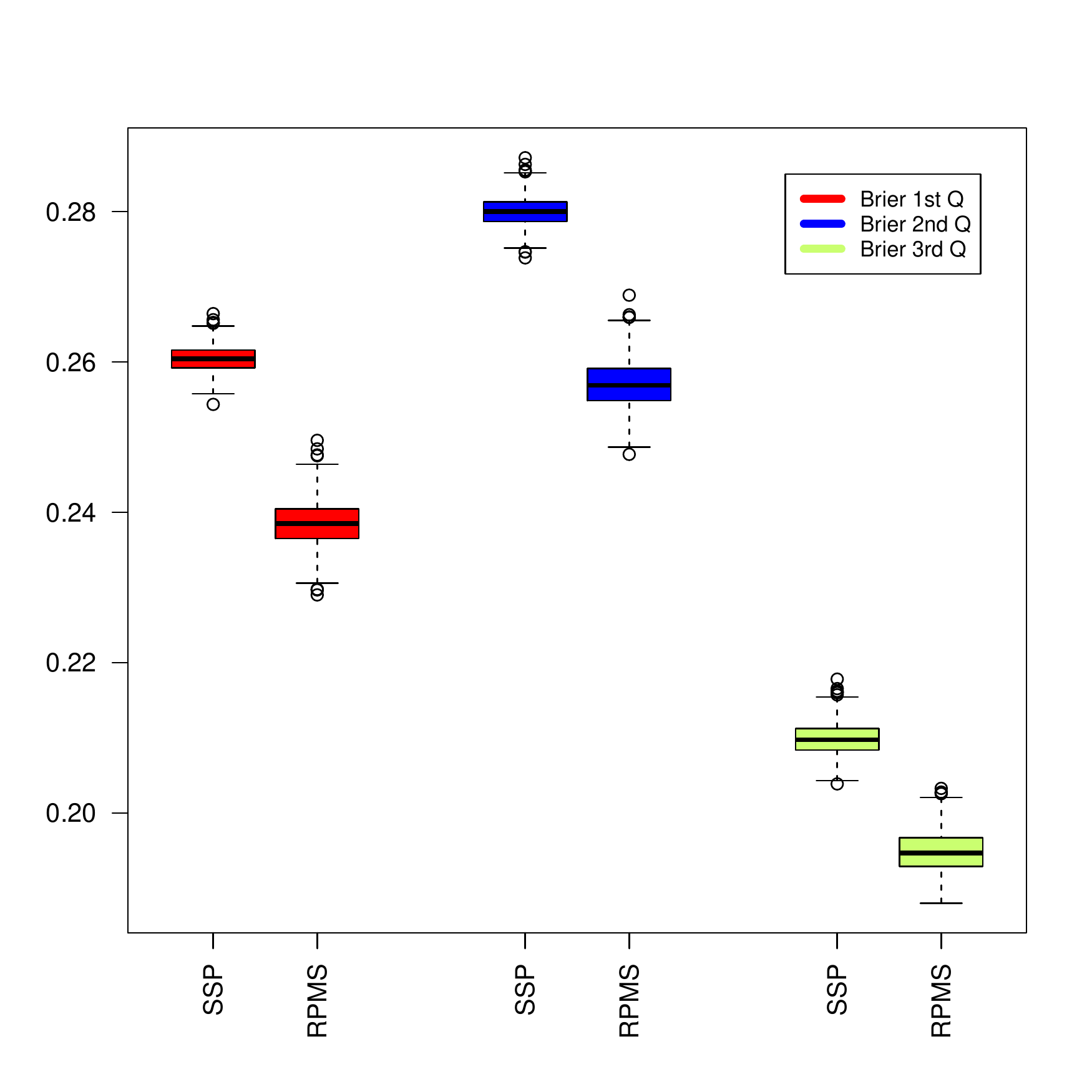}
\put(-435,+94){\rotatebox{90}{\small{Density}}}
\put(-347,+10){\small{$\log(\text{WBC})$}}
\caption{Left panel: kernel density estimate of the response variable $\log(\text{WBC})$. The vertical lines correspond to the quartiles. Right panel: posterior distributions of the Brier statistic using as threshold the first quartile, the second quartile (median) and third quartile. Results are shown for SSP and RPMS models.}
\label{fig:brier}
\end{center}
\end{figure}

We compare the posterior distribution of $\text{Brier}_{(q)}$ estimated by the SSP and the RPMS. Small values of the Brier statistic indicate good classification performance. The posterior distribution for the SSP and the RPMS are displayed in Figure \ref{fig:brier}, together with the kernel estimate of the response variable. We consider the first quartile, the median and the third quartile to discretise our response. In all three cases the location of the distribution of $\text{Brier}_{(q)}$ is lower for the RPMS model, indicating better predictive power compared to SSP.

Evidence in favour of the hypothesis that $\mbox{WBC} \geq 1$ indicates the presence of UTI is given in \cite{kupelianetal2013}. This recent result extends the analogous one of \cite{dukes1928}, where $\mbox{WBC} \geq 10$ was considered. All patients in our study have $\mbox{WBC} \geq 1$ and thus it is very likely they are affected by UTI. However, if on the one hand a large number of $\mbox{WBC}$ in a sample of urine will increase the confidence about the presence of UTI, on the other hand no work has been published that describes the severity of infection in relation to higher values of WBC. Nevertheless, specialists consider fully reasonable to associate  high degree of inflammation to large values of WBC. 

Hence, discretising the response variable to make prediction is reasonable both to assess the likelihood of having an infection and to evaluate the general status of the disease (in terms, for example, of the degree of inflammation). Moreover, discretising the response variable transforms the problem into a classification one, which links our model to the very important area of risk prediction models. A review of these models can be found in \cite{gerdsetal2008}, who highlight the most common techniques to perform model selection in this context.

\section{\textbf{Conclusions}}\label{sec:con}

In this work we have proposed the RPMS, a DPM of Normal regressions with covariate dependent weights, capable of simultaneously performing clustering and variable selection. This is achieved employing a DPM on the joint distribution of the response and the covariates, together with spike and slab prior distributions on the regression coefficients within the clusters of the partition. The latter allows performing variable selection and therefore identifying covariates with  high explanatory power on  the response. The proposed method is designed to handle binary covariates, due the dataset motivating this work, even though it is straightforward to  include other types of covariates (and also mixed types).  Although we have presented the model for Normally distributed response data,  it is possible to extend it to the generalised linear model framework. 

The main feature of the model lies in the fact that the RPMS takes into account also possible patterns within the covariate space. The results of the analysis highlight the diagnostic power of the symptoms. On the other hand, the SSP focuses on the variability within the response. The results of the posterior inference show that in the partition generated by the RPMS the clusters are characterized by the presence of certain classes of symptoms or  combinations of classes, revealing that these classes are informative and further investigation might lead to the identification of disease sub-types. 

The results of the variable selection has been summarized in two ways: fixing a meaningful partition (we have opted for the Binder estimate) or fixing a specific profile in a predictive fashion. In the first case, the analysis of the posterior distribution of the regression coefficients conditional to the Binder partition shows the overall importance of the urgency symptoms, and the cluster-specific importance of certain particular symptoms.  The second way to display the variable selection output is from a predictive perspective. We have assumed that a new patient's profile has been collected. The distribution of the regression coefficients for the new patient depends on the profile and this permits an individual-based assessment of the important symptoms that determine the distribution of WBC. The SSP's estimated posterior distributions of the regression coefficients instead do not depend on the patient profile. This difference permits the RPMS to achieve more accurate prediction of the WBC compared to the SSP.

We believe that the use of Bayesian non-parametric methods, although computationally more expensive, offers the flexibility necessary to capture the complexity of modern clinical data and consequently improved predictive power, especially in cases where the use of parametric approaches would impose unrealistic assumptions on the data generating process.

\section*{Supplementary Material}
The results of a simulation study are presented in Supplementary Material. 

\appendix
\section{\textbf{Posterior Inference}}\label{app:comp}

In this appendix we present the details of the updating steps of the Gibbs sampler scheme adopted.

\subsection*{Membership Indicator}
This step follows the updating Gibbs-type algorithm for DPM with non conjugate base measure in \cite{neal2000} called \textit{auxiliary parameter} approach. Let first define $s_i$ to be the membership indicator for the observation $i$ and $\bm{s}_{(i)}$ to be the vector of the membership indicator for the $n$ observations but from which $s_i$ is removed. Let us also define $k^-$ to be the number of clusters when $i$ is removed, $n_j^-$ for $j=1,\ldots, k^-$ to be the cardinality of the clusters when $i$ is removed. Thus the full conditional distribution for each indicator is:

\begin{dmath*}
p(s_i\mid \bm{s}_{(i)}, \bm{\beta}^*, \bm{\zeta}^*, \bm{X}, \bm{y}, \lambda) \propto \left\{\begin{array}{cc}
n_jp(y_i\mid \bm{x}_i, \bm{\beta}^*_j, \lambda)\prod_{d=1}^Dp(x_{id}\mid \zeta^*_{jd}) & j=1,\ldots,k^- \\
\frac{\alpha}{M}p(y_i\mid \bm{x}_i, \bm{\beta}^*_m, \lambda)\prod_{d=1}^Dp(x_{id}\mid \zeta^*_{md}) & j=k^-+1\text{ and } m=1,\ldots,M
\end{array} \right. 
\end{dmath*}
where $(\bm{\beta}^*_m,\bm{\zeta}^*_m)$ for $m=1,\ldots,M$ are draws from the the base measure in  (\ref{eq:g0}). 

\subsection*{Precision of the DP}
In order to update the parameter $\alpha$ of the DP we need to introduce an additional parameter $u$ such that $p(u\mid k, \alpha)=\text{Beta}(\alpha+1, n)$ (see \cite{escobaretal1995} for detailed explanation). We can sample from the full conditional:
\begin{displaymath}
p(\alpha\mid u, k) =  \xi\text{Gamma}(a_{\alpha}+k, b_{\alpha}-\log(u)) + (1-\xi)\text{Gamma}(a_{\alpha}+k-1, b_{\alpha}-\log(u))
\end{displaymath} 
where $\xi=(a_{\alpha}+k-1)/(\alpha_1+k-1+nb_{\alpha}-n\log(u))$.

\subsection*{Covariate Parameters}
For the update of the parameters of the covariates, we work separately for each of the $D$ covariates within each of the $k$ clusters. Thus the full conditional posterior distributions are:

\begin{displaymath}
p(\zeta_{jd}^*\mid \bm{x}_{jd}^*) \propto \prod_{i\in S_j} \text{Bernoulli}(x_{id}\mid \zeta_{jd}^*) \text{Beta}(\zeta_{jd}^*\mid a_{\zeta},b_{\zeta})
\end{displaymath}

\begin{displaymath}
\zeta_{jd}^*\mid \cdot \sim \text{Beta}\left(\zeta_{jd}^*\mid  a_{\zeta} + \sum_{i \in S_j}x_{id}, b_{\zeta} - \sum_{i \in S_j}x_{id} +n_j \right), \quad j=1,\ldots,k \quad \text{and} \quad d=1,\ldots,D
\end{displaymath}

\subsection*{Regression Coefficients}
As for the case of the parameters of the covariates, we consider separately each $D$ and each of the $k$ clusters. It follows that:

\begin{displaymath}
p(\beta_{jd}^*\mid \bm{X}, \bm{y}, \bm{\beta}_{j(d)}^*)\propto \prod_{i \in S_j} \mbox{Normal}(y_i\mid \bm{x}_i,\bm{\beta}_{j}^*,\lambda)[\pi_dw_{\omega}\delta_0(\beta_{jd}^*)+(1-\pi_dw_{\omega})\mbox{Normal}(\beta_{jd}^*\mid  m_d, \tau_d)]= 
\end{displaymath}

\begin{displaymath}
= \prod_{i \in S_j} \pi_dw_{\omega}\delta_0(\beta_{jd}^*)N(y_i\mid \bm{x}_i,\bm{\beta}_{j}^*,\lambda) + (1-\pi_dw_{\omega})\mbox{Normal}(\beta_{jd}^*\mid  m_d, \tau_d)\mbox{Normal}(y_i\mid \bm{x}_i,\bm{\beta}_{j}^*,\lambda) 
\end{displaymath}

$\bm{\beta}_{j(d)}^*$ is the vector $\bm{\beta}_{j}^*$ where the $d$th component is removed. The first part of the last equation will be 0 with some probability. Let us  consider the second part of that equation. This is proportional to
\begin{displaymath} 
\exp\left\{-\frac{1}{2}\tau_d(\beta_{jd}^*-m_d)^2\right\}\exp\left\{-\frac{1}{2}\sum_{i\in S_j}\lambda(y_i-\bm{x}_{i(d)}\bm{\beta}_{j(d)}^*-x_{id}\beta_{jd}^*)^2\right\}=
\end{displaymath}
\begin{displaymath}
=\exp\left\{-\frac{1}{2}\left[\beta_{jd}^{*2}(\tau_d+\lambda x_{id}^2)-2\beta_{jd}^*\left(m_d\tau_d+\lambda\sum_{i \in S_j}(x_{id}A_i)\right)\right]\right\}
\end{displaymath}
with $A_i=(y_i - \bm{x}_{i(d)}\bm{\beta}_{j(d)}^*)$, $\bm{x}_{i(d)}$ is the vector $\bm{x}_i$ where the $d$th is removed. 

Thus, the full conditional probabilities for the Gibbs sampler are:

\begin{displaymath}
\beta_{jd}^*\mid \cdot=\left\{\begin{array}{ll}
0& \textrm{w. p. $\theta_{jd}$ } \\
\sim \mbox{Normal}\left(\frac{m_d\tau_d + \sum_{i \in S_j}(\lambda x_{id}A_i)}{\tau_d + \sum_{i \in S_j}(\lambda x_{id}^{2})}, \tau_d + \sum_{i \in S_j}(\lambda x_{id})\right) & \textrm{w. p. $(1- \theta_{jd})$}
\end{array} \right.
\end{displaymath}

Finally the weights $\bm{\theta}=(\bm{\theta}_1,\ldots,\bm{\theta}_k)$ are:

\begin{displaymath}
\theta_{jd}=\frac{\pi_dw_{\omega}}{\pi_dw_{\omega}+(1-\pi_dw_{\omega})C}\end{displaymath}

with $C$:

\begin{dmath*}
C=\sqrt{\left(\tau_d+ \sum_{i \in S_j}(x_{id}^2\lambda)\right)^{-1}\tau_d}\exp\left\{-\frac{1}{2}\tau_dm_d^2 + \frac{1}{2}\left(\tau_d + \sum_{i \in S_j}(x_{id}^2\lambda)\right)^{-1}\cdot \left(m_d\tau_d + \sum_{i \in S_j}(\lambda x_{id}A_i)\right) \right\}
\end{dmath*}

\subsection*{Weights of the Spike and Slab Prior}
Following what was done in \cite{kimetal2009} let us call $r_d= \pi_dw_{\omega}$ and $w_{\omega}=a_{\omega}/(a_{\omega}+b_{\omega})$. 

\begin{displaymath}
p(r_d)= \text{Beta}\left(\frac{r_d}{w_{\omega}}\mid a_{\omega},b_{\omega}\right)\frac{1}{w_{\omega}}=
\end{displaymath}

\begin{displaymath}
\frac{1}{\text{B}(a_{\omega}, b_{\omega})}\left(\frac{1}{w_{\omega}}\right)^{a_{\omega}+b_{\omega}-1}r_d^{a_{\omega}-1}(w_{\omega}-r_d)^{b_{\omega}-1}
\end{displaymath}

The full conditional is the following:

\begin{displaymath}
p(r_d\mid \bm{\beta}_{d}^*)\propto p(r_d)r_d^{\sum_j1(\beta_{jd}^*=0)} (1-r_d)^{\sum_j1(\beta_{jd}^*\neq0)}
\end{displaymath}

This is an unknown distribution and a draw from it is obtainable computing the inverse of the cumulative distribution function over a grid of values. We select the point on the grid that gives the closest value of the inverse cumulative distribution function to a draw from a uniform distribution on $(0,1)$.

\subsection*{Precision of the Base Measure}
We will update the precision of the normal part of the base measure considering separately each of the $D$ covariates.

\begin{displaymath}
p(\tau_d\mid \bm{\beta}_j^*,a_{\tau},b_{\tau}) \propto \text{Gamma}(\tau_d\mid a_{\tau},b_{\tau})\prod_{j=1}^k[\pi_dw_{\omega}\delta_0(\beta^*_{jd})+(1-\pi_dw_{\omega})N(\beta^*_{jd}\mid m_d, \tau_d)]=
\end{displaymath}

\begin{displaymath}
=\text{Gamma}(\tau_d\mid a_{\tau},b_{\tau})\prod_{j=1}^{n_d^+}N(\beta_{jd}^+\mid m_d, \tau_d)
\end{displaymath}

where $n_d^+$ is the number of clusters that have non zero coefficients in position $d$, whereas $\beta_{jd}^+$ for $j=1,\ldots,n_d^+$ is the list of these non zero coefficients. Thus, it is possible to draw from the following known distribution:

\begin{displaymath}
\tau_d\mid  \bm{y}, \bm{\beta}_j^*,a_{\tau},b_{\tau}\sim \text{Gamma}\left(\tau_d\mid  a_{\tau} + \frac{n_d^{+}}{2} , b_{\tau}+ \frac{1}{2}\sum_{j=1}^{n_d^+}(\beta_{jd}^{+} - m_d)^2 \right)
\end{displaymath} 

\subsection*{Precision of the Regression}
The precision of the regression is updated in a conjugate form as it follows:

\begin{displaymath}
p(\lambda\mid \bm{y}, \bm{X}, \bm{\beta}^*,a_{\lambda}, b_{\lambda})\propto \prod_{j=1}^k\prod_{i\in S_j}\mbox{Normal}(y_i\mid \bm{x}_i, \bm{\beta}_j^*,\lambda)\text{Gamma}(\lambda\mid a_{\lambda}, b_{\lambda})
\end{displaymath}

\begin{displaymath}
\lambda\mid  \bm{y}, \bm{X}, \bm{\beta}^*,a_{\lambda}, b_{\lambda} \sim \text{Gamma}\left(\lambda\mid  n/2 + a_{\lambda}, \sum_{i=1}^n(y_i + \bm{x}_i\beta_i)^2/2 + b_{\lambda} \right)
\end{displaymath}

\bibliographystyle{ieeetr}
\bibliography{bibvarsel}
\end{document}